\documentclass[a4paper]{article}
\usepackage{graphicx}
\usepackage{epstopdf}
\usepackage{amsmath,amssymb,amsfonts}
\usepackage{a4wide,epic}
\usepackage{bm}
\usepackage{setspace}
\usepackage[T1]{fontenc}

\newcommand{\env}{\text{env}}
\newcommand{\BS}{\text{BS}}

\newcommand{\minou}{\text{-}}
\newcommand{\plusou}{\text{+}}

\newcommand{\ket}[1]{\left|#1\right>}
\newcommand{\bra}[1]{\left<#1\right|}
\newcommand{\tr}[1]{\text{Tr}\left(#1\right)}

\newcommand{\q}[1]{\vert #1 \rangle}
\newcommand{\qd}[1]{\langle #1 \vert}

\newcommand{\QNDg}{\xi}

\newcommand{\PP}{{\mathbb P}}
\newcommand{\EE}{{\mathbb E}}

\newcommand{\KK}{{\mathbb K}}
\newcommand{\NNN}{{\mathcal N}}
\newcommand{\CCC}{{\mathcal C}}

\begin{document}

\title{Loss-tolerant parity measurement for distant quantum bits}
\author{Alain Sarlette\thanks{QUANTIC project-team, INRIA Paris, France; and Department of Electronics and Information Systems, Ghent University, Belgium.}~~%
 and Mazyar Mirrahimi\thanks{QUANTIC project-team, INRIA Paris, France; and Department of Applied Physics, Yale University, USA}
}

\date{\today}

\maketitle

\begin{abstract}
We propose a scheme to measure the parity of two distant qubits, while ensuring that losses on the quantum channel between them does not destroy coherences within the parity subspaces. This capability enables deterministic preparation of highly entangled qubit states whose fidelity is not limited by the transmission loss. The key observation is that for a probe electromagnetic field in a particular quantum state, namely a superposition of two coherent states of opposite phases, the transmission loss stochastically applies a near-unitary back-action on the probe state. This leads to a parity measurement protocol where the main effect of the transmission losses is a decrease in the measurement strength. By repeating the non-destructive (weak) parity measurement, one achieves a high-fidelity entanglement in spite of a significant transmission loss.
\end{abstract}
\vspace{5mm}


The correlation of distant systems thanks to their entanglement is a proven hallmark of quantum physics \cite{bell:64,aspect-et-al:PRL81} and plays a fundamental role in envisioned quantum technology. Most fundamentally, quantum teleportation~\cite{Bennett-teleportation-93} shows how entanglement is a resource for effectively transmitting the unknown state of a quantum system between two locations, without physically transmitting quantum states. 
Towards the future quantum computer, such teleportation could transport information between the few-qubits processing units, and memory units which must be well isolated and hence should not be directly coupled to the rest of the system via physical interactions. {This so-called modular architecture for quantum computing provides a viable solution to the major scaling problem for many-qubit quantum information processing~\cite{Devoret-Schoelkopf-2013,Monroe-PRA-2014}.} In quantum communication, similar ideas would allow quantum repeaters to purify information through local operations only, provided they can consume units of entanglement between the two communicating devices~\cite{Briegel-PRL-98,Duan-Nature-2001}; such quantum repeaters are a necessary technology for exploiting accurate quantum communication, with associated e.g.~cryptographic benefits, over long distances. 

A major challenge towards enabling these applications is that generating entangled states between distant systems must rely, itself, on a quantum channel~\cite{Supp}. Microwave experiments have demonstrated how to deterministically entangle separate quantum subsystems via parity measurements \cite{Roch-Siddiqi-2014}, yet with a fidelity directly limited by the quality of the quantum channel: any losses on the probe field imply losses in entanglement. Channel losses can also be made to affect preparation success probability, instead of preparation fidelity. Indeed, many experiments in the optical domain have illustrated that by heralding the preparation on some rare photo-detection events, one can achieve significant entanglement despite propagation losses~\cite{Chou-Kimble-2005,Moehring-Monroe-2007,Ourjoumtsev-2009,Hofmann-Weinfurter-2012,Bernien-Hanson-2013}; a similar scheme with microwaves has recently been implemented in \cite{Narla-Devoret-2016}. The success rate of this probabilistic preparation is however usually very low. Furthermore, the preparation fidelity is still limited by imperfections such as the dark counts of the photodetector. 
The literature does not cover the possibility to highly entangle distant quantum bits \emph{deterministically} by using a lossy quantum channel. 

The present letter solves this question with an explicit proposal for the essentially equivalent \cite{Supp} achievement of an eigenstate-preserving quantum non-demolition (QND) parity measurement between spatially separated qubits. Although such parity measurement is not feasible with a quantum channel subject to arbitrary errors \cite{Supp}, it becomes solvable if the channel features one dominant error source. Hence our key idea is to transmit over the quantum channel particularly engineered quantum states of light, i.e.~``cat states'', for which the dominant photon loss errors almost reduce to photon-number parity flips \cite{Leghtas-al-PRL-2013}. With this we design the interaction between qubits and probe field such that (i) measuring the probe at the output performs a QND measurement of qubits parity and 
(ii) photon loss events on the transmitted probe field render the detection less decisive (weak measurement) but affect only minimally the parity eigenstates. 

\begin{figure}
\setlength{\unitlength}{1mm}
\textbf{a.} \begin{picture}(140,15)(0,1)
\put(2,8){$\q{\psi_p}$}
\put(10,9){\vector(1,0){6}}
\put(16,8){\framebox[9mm]{$\vphantom{\sum^A_B}\q{q_A}$}}
\put(17,2){$U_A$}
\put(25,9){\vector(1,0){35}}
\put(60,8){\framebox[9mm]{$\vphantom{\sum^A_B}\q{q_B}$}}\put(61,2){$U_B$}
\put(69,9){\vector(1,0){6}}
\qbezier(77,11)(75,9)(77,7)
\put(79,9){\line(-1,1){4}}
\put(79,9){\line(-1,-1){4}}
\put(79,9){\line(1,0){3}}
\put(40,11){$C$}
\qbezier(35,9)(37.5,12)(40,9)
\qbezier(40,9)(42.5,6)(45,9)
\put(28,4){losses/disturbances}
\end{picture}

\textbf{b.} \begin{picture}(110,30)(0,8)
\put(2,7){$\q{q_A}$}
\put(8,8){\line(1,0){60}}
\put(2,12){$\q{q_B}$}
\put(8,13){\line(1,0){60}}
\put(2,18){$\q{q_p}$}
\put(8,19){\line(1,0){15}}
\put(68,19){\line(-1,0){15}}
\qbezier(70,21)(68,19)(70,17)
\put(72,19){\line(-1,1){4}}
\put(72,19){\line(-1,-1){4}}
\put(72,19){\line(1,0){3}}
\put(26,24){\circle*{2}} \put(26,21.5){\line(0,1){3}}
\put(23,18){\framebox[7mm]{$E_1$}}
\put(49,27){\circle*{2}} \put(49,21){\line(0,1){6}}
\put(38,18){\framebox[7mm]{$E_k$}}
\put(41,26){\circle*{2}} \put(41,21.5){\line(0,1){5}}
\put(33,25){\circle*{2}} \put(33,21){\line(0,1){4}}
\put(32,19){...} \put(46,19){...}
\put(16,8){\circle*{2}}
\put(16,19){\circle{6}}\put(16,22){\line(0,-1){14}}
\put(60,13){\circle*{2}}
\put(60,19){\circle{6}}\put(60,22){\line(0,-1){9}}
\put(2,24){env}
\multiput(8,24)(0,1){4}{\line(1,0){60}}
\end{picture}
\vspace{10mm}

\textbf{c.} 
\includegraphics[width=100mm]{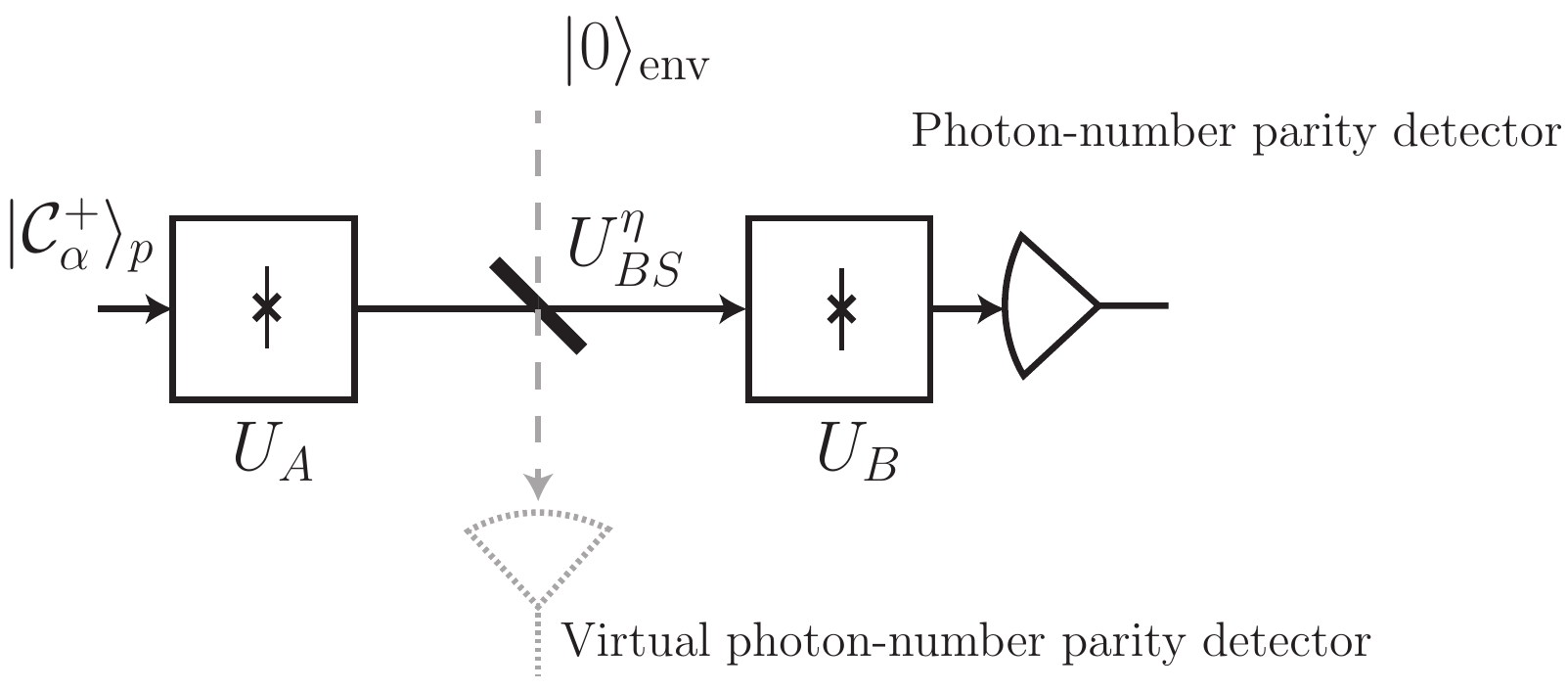}

\caption{\textbf{a.} General setup for remote parity measurement with a probe $\q{\psi_p}$ that propagates on a noisy quantum channel $C$ between the two target qubits $\q{q_A}$ and $\q{q_B}$. \textbf{b.} Quantum logic circuit summarizing our concept with CNOT gates involving target qubits $\q{q_A},\q{q_B}$ and a probe qubit $\q{q_p}$ that propagates along a quantum channel corrupted by unknown operations; we represent these as applying known $E_k$, $k=1,2,...$ conditionally on unknown states from the environment (env). \textbf{c.} Corresponding experimental setup with a probe field initialized in a coherent superposition of two opposite coherent states (``cat state''). \label{fig:scheme}}
\end{figure}

The abstract setting (Fig.\ref{fig:scheme}a) comprises two target qubits $\q{q_A}, \q{q_B}$ at different locations A, B and possibly embedded in auxiliary quantum machinery, e.g.~a cavity in circuit quantum electrodynamics (QED) setups~\cite{Wallraff-2004}. For each measurement, a source generates a controlled ``probe'' quantum state $\q{\psi_p}$ at A which then interacts with $\q{q_A}$ according to a unitary $U_A$, is transmitted over a noisy quantum channel $C$, before interacting with $\q{q_B}$ according to $U_B$ and finally hitting a detector at B. Those probe states play the role of parity meter. Since the quantum channel is the unequivocal bottleneck for remote entanglement in state-of-the-art technology~\cite{Lalumiere-Blais-2014}, we focus on this issue and assume in this letter that all (reasonable) local actions (i.e.~$U_A,U_B$, generating $\q{\psi_p}$, detection at $B$) are implemented perfectly.

The QND measurement of a quantum observable $Q$ discriminates possibly imperfectly between the eigenspaces of $Q$, but ensures that every \textit{eigenspace} of $Q$ remains unaffected for all possible detection results. {In the case of an observable $Q$ with degenerate eigenspaces, this only ensures that a state inside an eigenspace is sent to a state in the same eigenspace.} Here, we define a slightly stronger \textit{Eigenstate-Preserving Quantum Non-Demolition} (EP-QND) measurement, which stands for a QND measurement which does not affect any \textit{eigenstate} of the quantum observable $Q$. In other words the EP-QND property ensures that the measurement acts as identity on each eigenspace. Consider the parity observable $Q= Q_+ - Q_-$ associated to two qubits in the canonical basis $\{\q{0},\q{1}\}$, with
\begin{eqnarray*}
Q_+ = \q{00}\qd{00} + \q{11}\qd{11} & , &
Q_- = \q{01}\qd{01} + \q{10}\qd{10} \, .
\end{eqnarray*}
In a projective measurement of $Q$, a detection result $+$ (resp.$-$) would project the two qubits onto the even parity manifold $\text{span}\{\ket{00},\ket{11}\}$ (resp.~the odd parity manifold $\text{span}\{\ket{01},\ket{10}\}$). A less decisive EP-QND measurement can result for instance from classical uncertainty in the detection, e.g.~with probability $1\minou\QNDg$ an even (resp.~odd) parity state gives detection result $-$ (resp.~$+$). Then the probability to detect $+$ becomes
$p_+ = \QNDg \qd{\psi}  Q_+ \q{\psi} + (1\minou\QNDg) \qd{\psi} Q_- \q{\psi}$ with corresponding measurement back-action transforming initial state $\q{\psi}$ into~\cite{nielsen-chang-book}
$$\tilde{\KK}_+(\q{\psi}\qd{\psi}) = \frac{\QNDg Q_+ \q{\psi}\qd{\psi} Q_+  +  (1-\QNDg) Q_- \q{\psi}\qd{\psi} Q_-}{p_+} \; ,$$
and similarly for detection result $-$,
\begin{align*}
\tilde{\KK}_-(\q{\psi}\qd{\psi}) &= \frac{\QNDg Q_- \q{\psi}\qd{\psi} Q_-  +  (1-\QNDg) Q_+ \q{\psi}\qd{\psi} Q_+}{p_-} \\
p_- &= \QNDg \qd{\psi}  Q_- \q{\psi} + (1-\QNDg) \qd{\psi} Q_+ \q{\psi} \, .
\end{align*}
Any even-parity (resp.~odd-parity) state remains unchanged under such measurement back-action (whence the notation EP-QND) and predominantly gives detection result $+$ (resp.~$-$), i.e.~with probability $\QNDg \in (1/2,1]$. 
By repeating the EP-QND measurement sufficiently often, a precise conclusion about parity can be obtained without disturbing any initial state of definite parity.

We first sketch our concept with $\q{\psi_p} = \q{q_p}$ a probe qubit. Starting with $\q{q_p} = \q{0}_p$ we let $U_A$ (resp.$U_B$) implement a CNOT gate on $\q{q_p}$ conditioned by $\q{q_A}$ (resp.$\q{q_B}$), see Fig.\ref{fig:scheme}b. If the channel $C$ was perfect ($E_k=\text{Identity}$ for $k=1,2,...$ on Fig.\ref{fig:scheme}b), then an initial state $(\q{11} \pm \q{00})_{A,B}\, \q{0}_p\, /\sqrt{2}$ would remain unchanged, while an initial state $(\q{10} \pm \q{01})_{A,B}\, \q{0}_p\, /\sqrt{2}$ would come out as $(\q{10} \pm \q{01})_{A,B}\, \q{1}_p\, /\sqrt{2}$ just before detection of the probe. Thus the measurement operations correspond to $\tilde{\KK}_+,\tilde{\KK}_-$ with $\QNDg=1$.

Now let the channel subject the probe to an unknown number $n \in \{0,1,2,... \}$ of bit-flip operations $E_k = \q{1}\qd{0}_p + \q{0}\qd{1}_p$; this can be represented as a succession of CNOT gates, conditioning each bit-flip on an unknown hypothetical state of the environment. For $n$ even, the outcome is as for the perfect channel. For $n$ odd the final state of the probe is reversed, e.g.~input $(\q{11} \pm \q{00})_{A,B}\, \q{0}_p\, /\sqrt{2}$ yields output $(\q{11} \pm \q{00})_{A,B}\, \q{1}_p\, /\sqrt{2}$, but most importantly, the state of the target qubits remains unaffected. This essential property ensures that the expected evolution for $n$ unknown (equivalently, tracing over the unknown states of the environment) remains an EP-QND parity measurement, explicitly described by the operators $\tilde{\KK}_+,\tilde{\KK}_-$ with $\QNDg = \sum_{n\, \text{even}} Proba(n) < 1$. One can easily adapt this $\QNDg$ to account for detection misses and errors. A broad distribution of values of $n$ implies low contrast for the measurement, pushing $\QNDg$ close to 1/2, but it does not impede its EP-QND character. Hence when sufficiently many measurements can be repeated within a relevant timescale, a conclusive result is obtained even for $\QNDg$ very close to $1/2$ (see details below).

This EP-QND property is not retained under general channel errors. Indeed if for instance $C$ includes a phase-flip  $E_1=\q{0}\qd{0}-\q{1}\qd{1}_p$, then the initial even-parity Bell state $\q{\psi_+} = (\q{11}\plusou\q{00})_{A,B}/\sqrt{2}$ gets transformed by measurement back-action into $\q{\psi_-} = (\q{11}\minou\q{00})_{A,B}/\sqrt{2}$; thus this eigenstate of $Q$ is not conserved, breaking the EP-QND character. Not knowing if $E_1$ was applied or not, the initial pure entangled state $\q{\psi_+}$ gets transformed  into a statistical mixture of $\q{\psi_+}$ and $\q{\psi_-}$, i.e.~entanglement is lost. In fact it is impossible to ensure EP-QND measurement of parity when the channel disturbance can be arbitrary \cite{Supp}.

These abstract properties indicate a pathway towards loss-tolerant parity measurement: use a subspace of probe states on which the dominant decoherence channel acts as a unitary root of identity, e.g.~a bit-flip. This fits a physical implementation where the probe is an electromagnetic field pulse whose logical states are materialized by so-called ``cat states'', i.e. mesoscopic superpositions of coherent states.
This encoding, proposed earlier as a resource towards quantum computing  \cite{Leghtas-al-PRL-2013,Mirrahimi-NJP-2014}, ensures that successive photon losses imply in good approximation, coherent logical bit flips. The rest of this letter describes such implementation in detail.

This physical implementation of loss-tolerant parity measurement is sketched in Figure~\ref{fig:scheme}c. We denote
$$
\ket{\CCC_\beta^\pm}=(\ket{\beta}\pm\ket{-\beta}) \, / \, \NNN_\beta^{\pm} \,,\qquad \NNN_\beta^{\pm}=\sqrt{2\pm 2~ e^{-2|\beta|^2}},
$$
the superpositions between two coherent states $\ket{\beta}$ and $\ket{-\beta}$, $\beta \in \mathbb{R}$; the normalization constants $\NNN_\beta^\pm$ rapidly approach $\sqrt{2}$ as the coherent amplitude $\beta$ becomes large. The probe field is initially prepared in the state $\ket{\CCC_\alpha^+}_p$ and interacts with two qubit-cavity systems in a cascaded manner. Between the two setups, it is exposed to losses that are modeled by the mixing with the vacuum state $\ket{0}_{\env}$ of an ancillary mode. This is represented by a unitary operator $U_\BS^\eta$ modeling a beam-splitter Hamiltonian  with transmittance $\sqrt{\eta}$ and reflectance $\sqrt{1-\eta}$. The unitary operators apply:
\begin{align}\label{eq:unitary}
&U_A \ket{0}_{A}\ket{\CCC_\alpha^\pm}_p=\ket{0}_{A}\ket{\CCC_\alpha^\pm}_p, \qquad
U_A\ket{1}_{A}\ket{\CCC_\alpha^\pm}_p=\ket{1}_{A}\ket{\CCC_\alpha^\mp}_p,\\
&U_B \ket{0}_{B}\ket{\CCC_{\sqrt{\eta}\alpha}^\pm}_p=\ket{0}_{B}\ket{\CCC_{\sqrt{\eta}\alpha}^\pm}_p,\notag\qquad U_B\ket{1}_{B}\ket{\CCC_{\sqrt{\eta}\alpha}^\pm}_p=\ket{1}_{B}\ket{\CCC_{\sqrt{\eta}\alpha}^\mp}_p,\notag\\
&U_{\BS}^\eta\ket{\CCC_{\alpha}^\pm}_p\ket{0}_{\env}=\notag
\frac{1}{\NNN_\alpha^\pm} \left(\ket{\sqrt{\eta}\,\alpha}_p\ket{\sqrt{1\minou\eta}\,\alpha}_{\env}\pm\ket{\minou\sqrt{\eta}\,\alpha}_p\ket{\minou\sqrt{1\minou\eta}\,\alpha}_{\env}\right). \notag
\end{align}
Finally, after interaction with the second qubit, a measurement projects the probe's state onto $\ket{\CCC_{\sqrt{\eta}\alpha}^+}_p$ or $\ket{\CCC_{\sqrt{\eta}\alpha}^-}_p$. Identifying $\ket{\CCC_\beta^+}_p$ and $\ket{\CCC_\beta^-}_p$ respectively with the logical $\q{0}_p$ and $\q{1}_p$, we recover the above abstract scheme where $U_A,U_B$ implement CNOT gates. 

We now analyze the performance of this scheme.
Imagine a virtual detector for the ancillary field modeling the losses, also projecting it to one of the two states $\ket{\CCC_{\sqrt{1-\eta}\alpha}^+}_{\env}$ or $\ket{\CCC_{\sqrt{1-\eta}\alpha}^-}_{\env}$, but with unread detection result. The measurement outcomes of the two detectors (one real and one virtual) are associated to four Kraus operators $M_{\pm,\pm}$, modeling the back-action of the measurements on the target qubits: e.g. their state $\rho$, after measuring even parities for both detectors, should be modified to $M_{+,+}\rho M_{+,+}^\dag/\tr{M_{+,+}\rho M_{+,+}^\dag}$. Following the simple computations of the supplementary material, these Kraus operators are 
\begin{align}\label{eq:Kraus}
M_{+,+}&=\tfrac{\NNN_{\sqrt{1\minou\eta}\alpha}^+}{2}\left(\tfrac{\NNN_{\sqrt{\eta}\alpha}^+}{\NNN_\alpha^+}\ket{00}\bra{00}+\tfrac{\NNN_{\sqrt{\eta}\alpha}^-}{\NNN_\alpha^-}\ket{11}\bra{11}\right)\notag\\
M_{+,-}&=\tfrac{\NNN_{\sqrt{1\minou\eta}\alpha}^-}{2}\left(\tfrac{\NNN_{\sqrt{\eta}\alpha}^+}{\NNN_\alpha^-}\ket{10}\bra{10}+\tfrac{\NNN_{\sqrt{\eta}\alpha}^-}{\NNN_\alpha^+}\ket{01}\bra{01}\right)\notag\\
M_{-,+}&=\tfrac{\NNN_{\sqrt{1\minou\eta}\alpha}^+}{2}\left(\tfrac{\NNN_{\sqrt{\eta}\alpha}^-}{\NNN_\alpha^-}\ket{10}\bra{10}+\tfrac{\NNN_{\sqrt{\eta}\alpha}^+}{\NNN_\alpha^+}\ket{01}\bra{01}\right)\notag\\
M_{-,-}&=\tfrac{\NNN_{\sqrt{1\minou\eta}\alpha}^-}{2}\left(\tfrac{\NNN_{\sqrt{\eta}\alpha}^-}{\NNN_\alpha^+}\ket{00}\bra{00}+\tfrac{\NNN_{\sqrt{\eta}\alpha}^+}{\NNN_\alpha^-}\ket{11}\bra{11}\right).
\end{align}
Discarding the inaccessible outcome of the virtual detector, the back-action induced by the measurement of the probe field follows the partial Kraus maps:
\begin{align}\label{eq:dynamics}
\KK_{+}(\rho)&=\frac{M_{+,+}\rho M_{+,+}^\dag+M_{+,-}\rho M_{+,-}^\dag}{\tr{M_{+,+}\rho M_{+,+}^\dag+M_{+,-}\rho M_{+,-}^\dag}},\\ \nonumber
\KK_{-}(\rho)&=\frac{M_{-,+}\rho M_{-,+}^\dag+M_{-,-}\rho M_{-,-}^\dag}{\tr{M_{-,+}\rho M_{-,+}^\dag+M_{-,-}\rho M_{-,-}^\dag}}.
\end{align}
In the lossless case ($\eta=1$), $M_{+,-}$ and $M_{-,-}$ vanish as $\NNN^-_{0}=0$, and the coefficients in front of $\ket{00}\bra{00}$ and $\ket{11}\bra{11}$ in $M_{+,+}$ and in front of $\ket{01}\bra{01}$ and $\ket{10}\bra{10}$ in $M_{-,+}$ are identical, equal to 1. This corresponds to a projective parity measurement, as described above by $\tilde{\KK}_+,\tilde{\KK}_-$ with $\QNDg=1$.

The effect of transmission losses ($\eta<1$) is twofold. First, it reduces the measurement strength. Indeed, when the probe field is detected in a given parity (e.g.~$+$),
the qubits could be projected to the opposite parity manifold
(e.g.~by the Kraus operator $M_{+,-}$).
However, each measurement does increase the conditional probability of finding the qubits in the same parity manifold as the one indicated by the probe detections, because
$\NNN^+_{\sqrt{1-\eta}\alpha}\min\left(\frac{\NNN^+_{\sqrt{\eta}\alpha}}{\NNN^+_{\alpha}},\frac{\NNN^-_{\sqrt{\eta}\alpha}}{\NNN^-_{\alpha}}\right)>\NNN^-_{\sqrt{1-\eta}\alpha}\max\left(\frac{\NNN^+_{\sqrt{\eta}\alpha}}{\NNN^-_{\alpha}},\frac{\NNN^-_{\sqrt{\eta}\alpha}}{\NNN^+_{\alpha}}\right)$.
A projective parity measurement under perfect transmission ($\eta=1$) is thus replaced by a less decisive measurement for $\eta<1$, where at each shot we gain partial information on the parity. By repeating the measurement the state gets projected onto a well-defined parity subspace. An initial state of definite parity will always keep this parity (e.g.~$M_{+,-} (\q{00}+\q{11}) = 0$). 

The second, more harmful effect of the transmission loss is a slight perturbation of the EP-QND property, by introducing slow mixing within each given parity manifold. This is due to the coherent states $\ket{\beta}$ and $\ket{-\beta}$ 
not being perfectly orthogonal, so that 
$\NNN_\beta^+ \neq \NNN_\beta^-$. This effectively induces a dephasing inside parity manifolds, e.g.~$\frac{\NNN^+_{\sqrt{\eta}\alpha}}{\NNN^+_{\alpha}}>\frac{\NNN^-_{\sqrt{\eta}\alpha}}{\NNN^-_{\alpha}}$ implies that $M_{+,+}$ drives the even-parity states $\ket{B_{\pm}^e}=(\q{00}\pm\q{11})/\sqrt{2}$ towards $\q{00} = (\ket{B_+^e} + \ket{B_-^e})/\sqrt{2}$, while $M_{-,-}$ would drive them towards $\q{11} = (\ket{B_+^e} - \ket{B_-^e})/\sqrt{2}$.

\begin{figure}
\textbf{a.} \includegraphics[width=.42\linewidth]{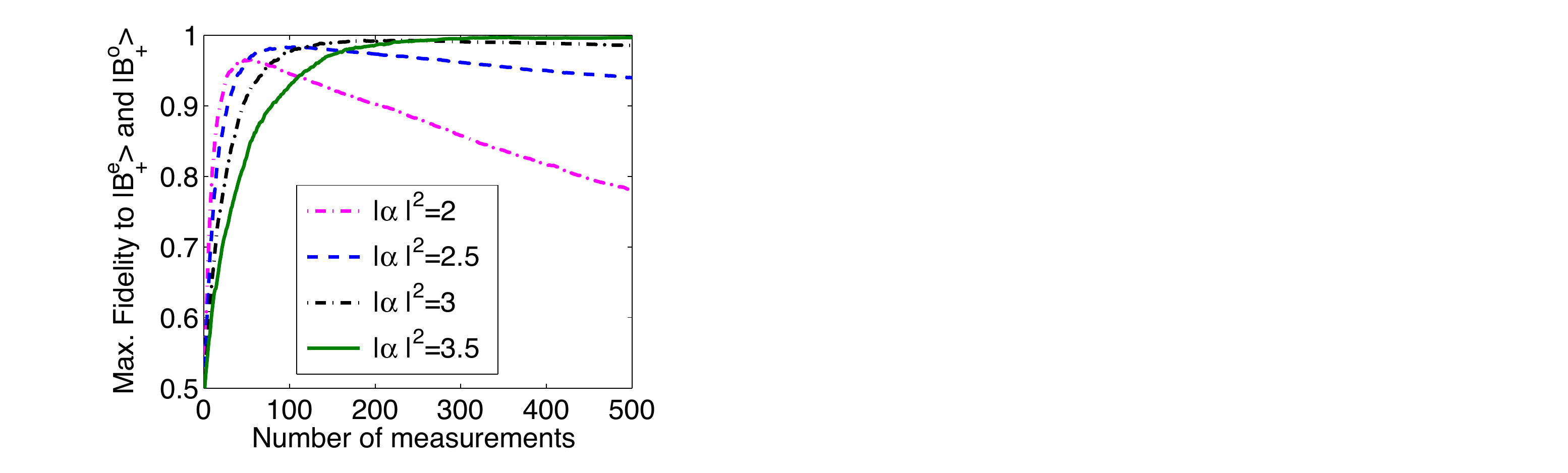} \hspace{10mm}
\textbf{b.} \includegraphics[width=.42\linewidth]{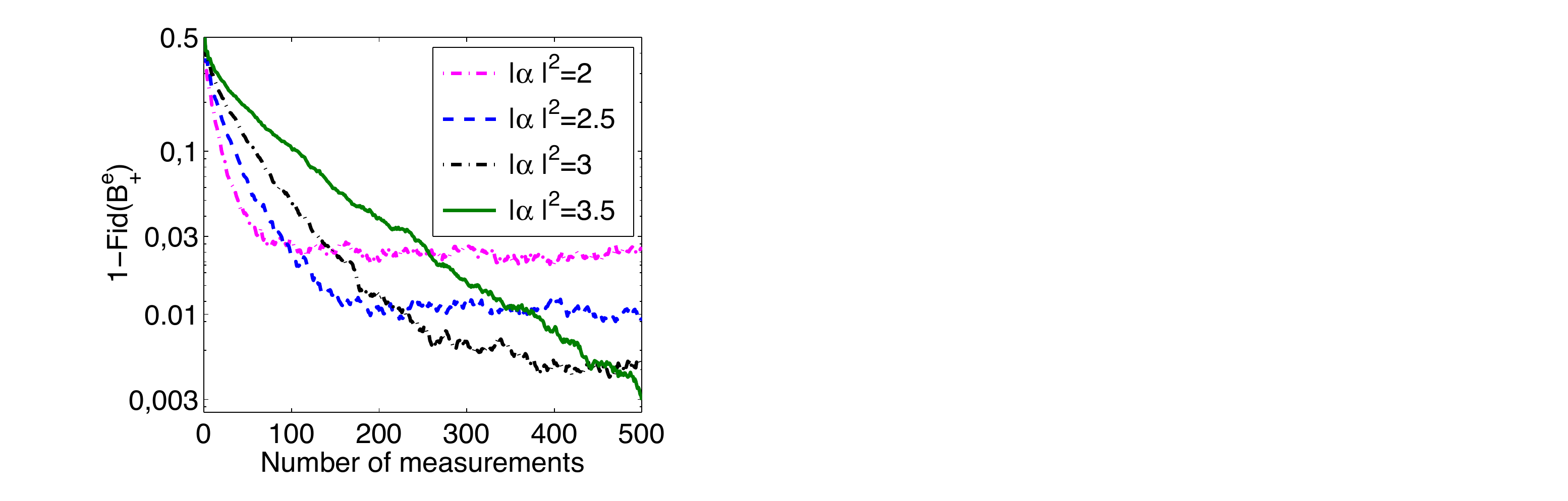}
\caption{\textbf{a.} Average evolution, over 1000 Monte-Carlo simulations, of the initial state $(\ket{0}+\ket{1})_A(\ket{0}+\ket{1})_B/2$ under repeated approximate EP-QND parity measurements \eqref{eq:Kraus},\eqref{eq:dynamics}, for various probe field intensities $\vert \alpha \vert^2$ and transmission efficiency $\eta=.75$. The state initially converges towards a definite parity state $\ket{B_+^e}=(\ket{00}+\ket{11})/\sqrt{2}$ or $\ket{B_+^o}=(\ket{01}+\ket{10})/\sqrt{2}$ thanks to the measurement, and then slowly loses fidelity to those states by dephasing, due to the slight perturbation of the EP-QND property inside definite parity manifolds. \textbf{b.} The dephasing can be counteracted by adding a simple feedback scheme (see main text), hence stabilizing $\ket{B_+^e}$ with high fidelity $Fid(B_e^+) = \bra{B_+^e} \rho \ket{B_+^e}$. \label{fig:FidVsNo}}
\end{figure}

The simulations of Fig.~\ref{fig:FidVsNo}a illustrate the competition between parity measurement and undesired dephasing while varying $|\alpha|^2$, the average number of photons in the probe field. Initializing both qubits in the state $\ket{+_X}=(\ket{0}+\ket{1})/\sqrt{2}$, an EP-QND parity measurement should project the joint state towards one of the two Bell states $\ket{B_+^e}=(\ket{00}+\ket{11})/\sqrt{2}$ or $\ket{B_+^o}=(\ket{01}+\ket{10})/\sqrt{2}$. This is the dominant tendency on Fig.~\ref{fig:FidVsNo}(a), while the transmission loss induces a slow dephasing mixing the target Bell states with the undesired ones $\ket{B_-^e}=(\ket{00}-\ket{11})/\sqrt{2}$ and $\ket{B_-^o}=(\ket{01}-\ket{10})/\sqrt{2}$. 
By increasing $|\alpha|^2$, this undesired dephasing gets suppressed significantly, at the expense of a slower convergence, i.e. weaker parity measurement. As soon as $\eta>1/2$, one can achieve arbitrarily high fidelity in this way.

This near EP-QND parity measurement can be used to stabilize a particular Bell state through a simple feedback protocol. For instance, to stabilize $\ket{B^e_+}$, (i) apply a $\pi$-pulse around the $X$-axis on the first qubit whenever the measurements estimate a probability higher than $1/2$ to be in the odd parity manifold, (ii) after that, apply a $\pi/2$-pulse on both qubits around the $Y$-axis irrespectively of the detection result. The measurement back-action favors convergence towards the dominant parity, the $\pi$ pulse correcting the parity whenever the state is converging towards the wrong one. This pushes the state towards the span of $\ket{B_+^e}$,$\ket{B_-^e}$ without favoring the target $\ket{B_+^e}$. The two $\pi/2$-pulses then leave $\ket{B_+^e}$ untouched and send the undesired $\ket{B_-^e}$ onto $\ket{B_+^o}$, such that the next parity measurement stochastically moves the corresponding population as well towards the target Bell state. The simulations of Fig.~\ref{fig:FidVsNo}b illustrate the performance of this protocol, having fixed $\eta=.75$ and varying $|\alpha|^2.$ Feedback stabilization of entanglement is further discussed in the supplementary material.

\begin{figure}
\includegraphics[height=80mm, trim=0mm 0mm 0mm 0mm, clip=true]{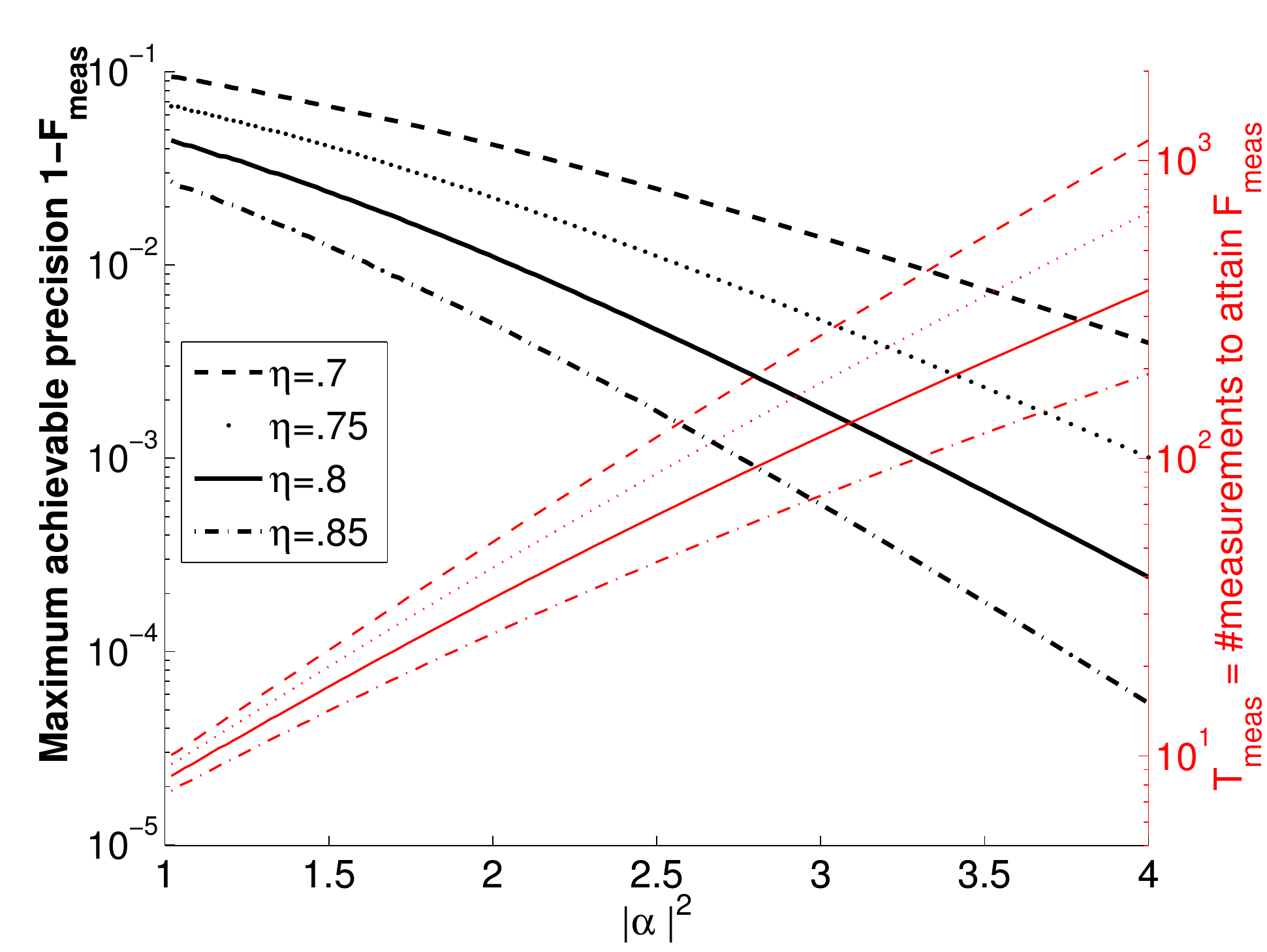}
\caption{{(thick black, left axis) Estimate $F_{\text{meas}}$ of the highest fidelity to the closest Bell state, obtained by repeated application of our near EP-QND parity measurement \eqref{eq:Kraus},\eqref{eq:dynamics} starting from $(\ket{0}+\ket{1})_A(\ket{0}+\ket{1})_B/2$, before dephasing destroys its coherence in absence of feedback; (thin red, right axis) Estimate $T_{\text{meas}}$ of the number of measurements after which this highest fidelity is reached, giving an indication of the measurement strength.}\label{fig:FidVsTime}}
\end{figure}

Convergence rates can be calculated analytically for both the parity measurement and the spurious dephasing, yielding respectively \cite{Supp}:
\begin{align*}
r_{\text{parity}}&=\frac{1}{2}\log\left(\frac{1-e^{-4|\alpha|^2}}{1-e^{-4(1-\eta)|\alpha|^2}}\right) \, ,\\ 
r_{\text{dephasing}}&=\frac{1}{2}\log\left(\frac{1-e^{-4|\alpha|^2}}{1-e^{-4\eta|\alpha|^2}}\right).
\end{align*}
This allows to {estimate the  measurement performance} as a function of $\eta$ and $|\alpha|^2$. Consider again the evolution depicted on Fig.\ref{fig:FidVsNo}a. The fidelity to the closest Bell state is 1/2 times the sum of two terms: dominant parity population, which converges from $1/2$ to $1$ at {roughly} a rate $r_\text{parity}$, and dominant phase population, which decreases from $1$ to $1/2$ at a rate $r_\text{dephasing}$. The two terms contribute equally to the error after a number of measurements $T=T_{\text{meas}}$ that satisfies $e^{-r_{\text{parity}}T}+e^{-r_{\text{dephasing}} T}=1$. The corresponding estimate of Bell state fidelity is $F_{\text{meas}} = 1-e^{-r_{\text{parity}}T_{\text{meas}}}/2$. By solving {numerically} the above transcendental equation, Figure \ref{fig:FidVsTime} illustrates these estimates of our  parity measurement's performance. For a transmission efficiency as low as $70\%$ we can get $F_{\text{meas}}$ as high as $99\%$ with less than 400 measurement runs, taking $\vert \alpha \vert^2 = 3.273$. Increasing $\eta$ to $85\%$ one can achieve the same $F_{\text{meas}}$ with only 17 measurements, taking $\vert \alpha \vert^2 = 1.63$.

All the required operations for this proposal have been individually implemented within the framework of quantum superconducting circuits. The strong dispersive coupling of a transmon qubit to a high-Q cavity mode~\cite{schuster-nature07}, provides the universal controllability of the state of the quantum harmonic oscillator modeling the cavity mode~\cite{Leghtas-al-PRA-2013,Krastanov-Jiang-2015}.  This controllability has been experimentally illustrated with circuit QED setups~\cite{Vlastakis-et-al-Science_2013,Heeres-Schoelkopf-2015}. Such a coupling enables to prepare the probe field in a cat state and to perform the CNOT gates $U_A$ and $U_B$ of~\eqref{eq:unitary} between the qubits $A$ and $B$ and associated intra-cavity fields. Recent experiments realizing a variable coupling between cavity modes and a transmission line~\cite{Yin-Martinis-2013,Wenner-Martinis-2014,Flurin-Huard-2015}, provide the possibility of catching the propagating microwave field, performing the required gate between the qubit and the cavity field, and finally releasing back the cavity photons. Finally, while the measurement of photon-number parity has also been realized in a similar setup~\cite{Sun-Schoelkopf-2013}, one can further simplify the protocol by {letting $U_B$ map the intra-cavity field to coherent states $\ket{\pm\sqrt{\eta}\alpha}$ instead of the cat states $\ket{\CCC_{\sqrt{\eta}\alpha}^\pm}$.}
This would allow to replace a photon-number parity measurement by a simple homodyne detection of the released field using a parametric amplifier~\cite{Vijay-Devoret-Siddiqi-2009}. A single measurement duration in such an experimental realization depends mainly on the time required to perform the operations $U_A$ and $U_B$. As explained in~\cite{Leghtas-al-PRA-2013,Krastanov-Jiang-2015} this gate time is roughly the inverse of the dispersive coupling strength. Recent experiments where this coupling strength is more than three orders of magnitude larger than both the qubit and the cavity decay rates~\cite{Vlastakis-et-al-Science_2013,Paik-et-al-PRL2011,Heeres-Schoelkopf-2015} indicate that high entanglement fidelities should be achievable whenever a transmission efficiency of more than $70\%$ is achieved. 

We have shown that it is possible to perform a near EP-QND measurement of the parity of two distant qubits despite an important loss through the transmission channel between them. By preparing the probe field in a quantum superposition of two coherent states with opposite phases, we avoid any back-action of the probe losses on the target qubits' state inside any parity subspace. Indeed, such losses mainly decrease the measurement strength but barely affect its EP-QND character. Therefore, even with an inefficient transmission channel, by repeating the measurement many times one can efficiently project the joint qubits state onto a definite parity subspace. This could be used not only to deterministically and efficiently prepare an entangled state of two distant qubits, but combined with a quantum feedback strategy it could even protect such entangled state against the local decay channels of the qubits \cite{Supp}. The operations required to perform this loss-tolerant parity measurement are within the reach of state of the art experiments with quantum superconducting circuits. Their implementation will lead to an important step forward {for implementing quantum teleportation protocols in a loss-tolerant way, and more particularly} towards the modular architecture solution for large scale quantum information processors~\cite{Devoret-Schoelkopf-2013}.\\

The authors thank the Agence Nationale de la Recherche for financial support under grant ANR-14-CE26-0018.



\newpage

\section*{\uppercase{Supplementary material}}

\subsection*{Constraining the channel loss operators is necessary for QND parity measurement}

We prove that it is impossible to design the measurement setup of Fig.1a of the main text in order to obtain an EP-QND parity measurement under arbitrary channel noise, i.e.~when the $E_k$ in Fig.1b of the main text can be arbitrary. 

\begin{itemize}
\item[S0.] Note that any unconditioned local unitaries on the qubits along the measurement chain, can be merged into $U_A$ and $U_B$. Indeed, even if physically we apply such $\tilde{U}_A$ to qubit $A$ after $U_B$, this $\tilde{U}_A$ commutes with $U_B$ since it acts as identity on both $\q{q_B}$ and $\q{\psi}_p$. We will thus assume without loss of generality a setting with just two arbitrary unitaries $U_A$ and $U_B$, which make the arbitrary probe interact respectively with $\q{q_A}$ and $\q{q_B}$.

\item[S1.] In order to keep invariant the even-parity states $\q{00}_{A,B}$ and $\q{11}_{A,B}$, we have to take $U_A$ such that $U_A \q{0}_A\q{\psi_p} = \q{0}_A\q{\phi_0}_p$ and $U_A \q{1}_A\q{\psi_p} = \q{1}_A\q{\phi_1}_p$ for some $\q{\phi_0},\q{\phi_1}$. Indeed consider \emph{a contrario} that e.g.~an initial $\q{0}_A$ was mapped by $U_A$ into a state involving $\q{1}_A$; then, since no further action is applied on qubit $A$, at the end of the nominal measurement chain there would unavoidably be a nonzero probability to end up in a state involving this $\q{1}_A$, which would contradict the EP-QND objective for an initial state $\q{00}_{A,B}$.

\item[S2.] If we chose $\q{\phi_1} = e^{i\theta} \q{\phi_0}$ for some $\theta$, then we could rewrite
$$U_A \q{q_A} \q{\psi_p}= (Z \q{q_A}) \q{\phi_0}_p$$
with $Z$ acting only on qubit $A$ and defined by $Z\q{0}_A = \q{0}_A$, $Z\q{1}_A = e^{i\theta} \q{1}_A$. This clarifies that the probe state $\q{\phi_0}_p$ would carry no information about the state of qubit $A$, hence we can get no parity information by later measuring the probe system at $B$.

\item[S3a] On the other hand if we have $\q{\phi_1} \neq e^{i\theta}\, \q{\phi_0}$, then it is always possible to write a noise channel which keeps $\q{\phi_0}$ in place but, conditionally on the unknown state $\q{q_\env}$ of some environment variable, moves $\q{\phi_1}$. I.e.~referring to Fig.1b of the main text, 
$$E_1\,\q{\phi_0}_p = \q{\phi_0}_p  \quad \text{and} \quad E_1\, \q{\phi_1}_p = \q{\phi_2}_p \neq \q{\phi_1}_p\, .$$

\item[S3b] Consider now what happens to the initial even-parity state $(\q{00}+\q{11})_{A,B}/\sqrt{2}$. After $U_A$ has been applied, the qubits are entangled with the probe in the state $(\q{00}_{A,B}\q{\phi_0}_p+\q{11}_{A,B}\q{\phi_1}_p) / \sqrt{2}$. In order to have an EP-QND measurement, the remaining action $U_B$ has to disentangle $(\q{00}+\q{11})_{A,B}/\sqrt{2}$ from the probe before detection. In the case where $E_1$ is not applied (no channel loss), we thus need
\begin{equation}\label{eq:eb}
U_B \q{0}_B\q{\phi_0}_p = \q{0}_B\q{\phi_b}_p \; \text{\scriptsize AND }  U_B \q{1}_B\q{\phi_1}_p = \q{1}_B\q{\phi_b}_p
\end{equation}
for some $\q{\phi_b}$. In the case where $E_1$ in contrast is applied by the lossy channel, we need
\begin{equation}\label{eq:ea}
U_B \q{0}_B\q{\phi_0}_p = \q{0}_B\q{\phi_a}_p \; \text{\scriptsize AND } U_B \q{1}_B\q{\phi_2}_p = \q{1}_B\q{\phi_a}_p
\end{equation}
for some $\q{\phi_a}$. For the experiment we must select a unique $U_B$, without knowing the environment state $\q{q_\env}$ i.e.~without knowing whether $E_1$ was applied or not. The left parts of \eqref{eq:eb},\eqref{eq:ea} thus impose 
$$\q{\phi_b} = \q{\phi_a} \; .$$
But then the second parts of \eqref{eq:eb},\eqref{eq:ea} lead to a contradiction, as they require the unitary $U_B$ to map two different initial states $\q{1}_B\q{\phi_1}_p \neq \q{1}_B\q{\phi_2}_p$ onto the same final state.
\end{itemize}
This shows the impossibility for the measurement to be EP-QND with respect to the ``conditional $E_1$'' channel noise. We have taken care to keep our construction fully general, such that we can conclude: whatever our design for $\q{\psi_p}$, $U_A$, $U_B$, we will obtain a measurement setup either which gains no parity information, or for which some particular noise actions $E_1$ can destroy the EP-QND character and perturb definite-parity states.\\ 

For example, in our proposed construction with CNOT gates: 
\begin{itemize}
\item[-] For S1 we have $\q{\phi_0}=\q{0}$, $\q{\phi_1}=\q{1}$; we are obviously in the situation of S3a, not of S2.
\item[-] Regarding S3a, the phase-flip channel $\sigma_z$ for instance would keep $\q{0}$ at $\q{0}$ but move $\q{1}$ to $-\q{1}$. Then $U_B$ to save the situation would have to satisfy in particular
$U_B \q{1}_B\q{1}_p = -U_B \q{1}_B\q{1}_p \; ,
$
 which is not possible.
\end{itemize}
This example shows that an EP-QND parity measurement is not possible if the transmitted probe state can be subject to both bit-flip and phase-flip errors.

\subsection*{Channel with constrained loss operators: generalization of loss-tolerant measurement}

In the main text we discuss the case, motivated by realistic experimental conditions, where the transmission channel only subjects the probe to an unknown number of bit-flips. This is not the only favorable situation.

Consider indeed a situation where the transmission channel subjects the probe to an unknown number $n$ of unitary operations of the same type $U_C$, where $(U_C)^N = $ Identity for some integer $N$. Now let $U_A$ and $U_B$ apply conditioned $V$-gates on the probe, where 
$$V = (U_C)^{N/2} \; .$$ 
Then all the operations $U_A$, $U_B$, $U_C$ commute with each other, and an initial probe state $\q{\psi_0}_p$ gets mapped just before detection onto 
$$(U_C)^{N/2}\cdot (U_C)^{N/2} \cdot U_C^n \, \q{\psi_0}_p = U_C^n \, \q{\psi_0}_p$$
if the target qubits have even parity, or onto 
$$(U_C)^{N/2} \cdot U_C^n \, \q{\psi_0}_p$$
if they have odd parity, while the target qubits remain unaffected. Thus not knowing $n$ makes the final probe state uncertain, as with the bit-flip, implying that detection results will not allow to perfectly discriminate the parity. But the probe ends up in the \emph{same} (unknown) state for all even parity states of the target qubit; and it ends up in another same state for all odd parity states of the target qubit. This ensures preservation of the EP-QND property under channel losses.

\subsection*{Kraus operators computation}

We here summarize the calculations that lead to the Kraus operators of~Eq.(2) in the main text. We start with an initial joint state of the two qubits, the probe field, and the ancillary mode modeling the transmission loss, given by
$$
\ket{\psi_0}=\big(c_{00}\ket{00}_{A,B}+c_{11}\ket{11}_{A,B}+c_{01}\ket{01}_{A,B}+c_{10}\ket{10}_{A,B}\big)\ket{\CCC_\alpha^+}_p\ket{0}_{\env}.
$$
Following the  definition of the unitary operators $U_A$, $U_B$ and $U_\BS^\eta$ in~Eq.(1) of the main text, just before performing the photon-number parity measurements of the probe and ancillary fields, this joint state has evolved to:
\begin{align*}
\ket{\psi}&=\frac{c_{00}}{2\,\NNN_\alpha^+}\ket{00}_{A,B}\Big[
\ket{\CCC_{\sqrt{\eta}\alpha}^+}_p \ket{\CCC_{\sqrt{1-\eta}\alpha}^+}_\env \NNN_{\sqrt{\eta}\alpha}^+ \NNN_{\sqrt{1-\eta}\alpha}^+
+ \ket{\CCC_{\sqrt{\eta}\alpha}^-}_p \ket{\CCC_{\sqrt{1-\eta}\alpha}^-}_\env \NNN_{\sqrt{\eta}\alpha}^- \NNN_{\sqrt{1-\eta}\alpha}^-
\Big] \\
&+\frac{c_{11}}{2\,\NNN_\alpha^-}\ket{11}_{A,B}\Big[
\ket{\CCC_{\sqrt{\eta}\alpha}^+}_p \ket{\CCC_{\sqrt{1-\eta}\alpha}^+}_\env \NNN_{\sqrt{\eta}\alpha}^- \NNN_{\sqrt{1-\eta}\alpha}^+
+ \ket{\CCC_{\sqrt{\eta}\alpha}^-}_p \ket{\CCC_{\sqrt{1-\eta}\alpha}^-}_\env \NNN_{\sqrt{\eta}\alpha}^+ \NNN_{\sqrt{1-\eta}\alpha}^-
\Big] \\
&+\frac{c_{01}}{2\NNN_\alpha^+}\ket{01}_{A,B}\Big[
\ket{\CCC_{\sqrt{\eta}\alpha}^+}_p \ket{\CCC_{\sqrt{1-\eta}\alpha}^-}_\env \NNN_{\sqrt{\eta}\alpha}^- \NNN_{\sqrt{1-\eta}\alpha}^-
+ \ket{\CCC_{\sqrt{\eta}\alpha}^-}_p \ket{\CCC_{\sqrt{1-\eta}\alpha}^+}_\env \NNN_{\sqrt{\eta}\alpha}^+ \NNN_{\sqrt{1-\eta}\alpha}^+
\Big] \\
&+\frac{c_{10}}{2\NNN_\alpha^-}\ket{10}_{A,B}\Big[
\ket{\CCC_{\sqrt{\eta}\alpha}^+}_p \ket{\CCC_{\sqrt{1-\eta}\alpha}^-}_\env \NNN_{\sqrt{\eta}\alpha}^+ \NNN_{\sqrt{1-\eta}\alpha}^-
+ \ket{\CCC_{\sqrt{\eta}\alpha}^-}_p \ket{\CCC_{\sqrt{1-\eta}\alpha}^+}_\env \NNN_{\sqrt{\eta}\alpha}^- \NNN_{\sqrt{1-\eta}\alpha}^+
\Big].
\end{align*}
Here, we have used the fact that
$$
\ket{\pm\beta}=\frac{\NNN_{\beta}^+}{2}\ket{\CCC_\beta^+}\pm\frac{\NNN_{\beta}^-}{2}\ket{\CCC_\beta^-} \; ,$$
for $\beta=\sqrt{\eta}\alpha$ and $\beta=\sqrt{1-\eta}\alpha$.
Detecting the probe and the ancillary fields both in even parity leads to applying the following projection operator as the measurement back-action
\begin{eqnarray*}
\Pi_{+,+}&=&\text{Id}_{A,B}\otimes \Pi^{\text{even}}_{p} \otimes\Pi^{\text{even}}_{\env}\\
&=&\text{Id}_{A,B} \otimes(\sum_{k=0}^\infty \ket{2k}\bra{2k}_p)\otimes (\sum_{k=0}^\infty \ket{2k}\bra{2k}_\env)
\end{eqnarray*}
with $\text{Id}$ the identity map. Since $\ket{\CCC_\beta^+}$ has even parity and $\ket{\CCC_\beta^-}$ has odd parity, the projected wave function can be simply read off the above expression,
\begin{align*}
\Pi_{+,+}\ket{\psi}=\frac{\NNN_{\sqrt{1\minou\eta}\alpha}^+}{2}&\left(c_{00}\frac{\NNN_{\sqrt{\eta}\alpha}^+}{\NNN_\alpha^+}\ket{00}+c_{11}\frac{\NNN_{\sqrt{\eta}\alpha}^-}{\NNN_\alpha^-}\ket{11}\right)_{A,B} \;\ket{\CCC_{\sqrt{\eta}\alpha}^+}_p\ket{\CCC_{\sqrt{1-\eta}\alpha}^+}_{\env}.
\end{align*}
In a similar way, we have the following projected wave functions for other measurement outcomes
\begin{align*}
\Pi_{+,-}\ket{\psi}=\frac{\NNN_{\sqrt{1\minou\eta}\alpha}^-}{2}&\left(c_{01}\frac{\NNN_{\sqrt{\eta}\alpha}^-}{\NNN_\alpha^+}\ket{01}+c_{10}\frac{\NNN_{\sqrt{\eta}\alpha}^+}{\NNN_\alpha^-}\ket{10}\right)_{A,B} \;
\ket{\CCC_{\sqrt{\eta}\alpha}^+}_p\ket{\CCC_{\sqrt{1-\eta}\alpha}^-}_{\env}
\end{align*}
\begin{align*}
\Pi_{-,+}\ket{\psi}=\frac{\NNN_{\sqrt{1\minou\eta}\alpha}^+}{2}&\left(c_{01}\frac{\NNN_{\sqrt{\eta}\alpha}^+}{\NNN_\alpha^+}\ket{01}+c_{10}\frac{\NNN_{\sqrt{\eta}\alpha}^-}{\NNN_\alpha^-}\ket{10}\right)_{A,B} \;
\ket{\CCC_{\sqrt{\eta}\alpha}^-}_p\ket{\CCC_{\sqrt{1-\eta}\alpha}^+}_{\env}
\end{align*}
\begin{align*}
\Pi_{-,-}\ket{\psi}=\frac{\NNN_{\sqrt{1\minou\eta}\alpha}^-}{2}&\left(c_{00}\frac{\NNN_{\sqrt{\eta}\alpha}^-}{\NNN_\alpha^+}\ket{00}+c_{11}\frac{\NNN_{\sqrt{\eta}\alpha}^+}{\NNN_\alpha^-}\ket{11}\right)_{A,B} \;
\ket{\CCC_{\sqrt{\eta}\alpha}^-}_p\ket{\CCC_{\sqrt{1-\eta}\alpha}^-}_{\env}.
\end{align*}
This corresponds to the Kraus operators as defined by~Eq.(2) in the main text.

\subsection*{Convergence rates}
The rate at which the parity measurement acquires information can be expressed analytically as a function of $\eta$ and $|\alpha|^2$. {Define the Lyapunov function
\begin{eqnarray}
\label{eq:Lyapa}
V^{\text{parity}}(\rho)& =& \sqrt{\bra{00}\rho\ket{00}\bra{10}\rho\ket{10}}\\
\nonumber & &  +\sqrt{\bra{11}\rho\ket{11}\bra{01}\rho\ket{01}} \\[3mm]
\nonumber & = & \frac{w_{B0}}{2} \sqrt{1-(P_{B0})^2} + \frac{w_{B1}}{2} \sqrt{1-(P_{B1})^2} \;\; ,
\end{eqnarray}
where $w_{Bq_B} = \bra{q_B} \rho \ket{q_B}_B$ denotes the population with qubit $B$ in state $\ket{q_B}$ and 
$P_{Bq_B} = \tr{\sigma_z\otimes\sigma_z \; \rho \; \ket{q_B}\bra{q_B}} / w_{Bq_B}$ measures the parity, conditioned on qubit $B$ being in state $\ket{q_B}$, for $q_B \in \{0,1\}$.}
One can show that
\begin{equation}\label{eq:Lyap}
\langle V^{\text{parity}}(\rho_{k+1})\rangle=\frac{\sqrt{1-e^{-4(1-\eta)|\alpha|^2}}}{\sqrt{1-e^{-4|\alpha|^2}}}\langle V^{\text{parity}}(\rho_{k})\rangle,
\end{equation}
where $\rho_k$ is the joint qubits state after the $k$'th measurement and $\langle V^{\text{parity}} \rangle$ denotes the ensemble average of $V^{\text{parity}}$ over measurement realizations. Indeed, we have
$$
\EE(V^{\text{parity}}(\rho_{k+1})~|~\rho_k)= V^{\text{parity}}(\KK_+(\rho_k))\PP(+~|~\rho_k)+V^{\text{parity}}(\KK_-(\rho_k))\PP(-~|~\rho_k),
$$
where 
\begin{align*}
\PP(+~|~\rho_k)&=\tr{M_{+,+}\rho_k M_{+,+}^\dag+M_{+,-}\rho_k M_{+,-}^\dag}\\
\PP(-~|~\rho_k)&=\tr{M_{-,+}\rho_k M_{-,+}^\dag+M_{-,-}\rho_k M_{-,-}^\dag}
\end{align*}
are respectively the conditional probabilities of achieving a positive/negative  outcome at $k$'th measurement. Therefore, we have
\begin{align*}
\EE(V^{\text{parity}}(\rho_{k+1})~|~\rho_k)= &\sqrt{\bra{00}M_{+,+}\rho_k M_{+,+}^\dag \ket{00}\bra{10}M_{+,-}\rho_k M_{+,-}^\dag \ket{10}}+\\
&\sqrt{\bra{11}M_{+,+}\rho_k M_{+,+}^\dag \ket{11}\bra{01}M_{+,-}\rho_k M_{+,-}^\dag \ket{01}}+\\
&\sqrt{\bra{00}M_{-,-}\rho_k M_{-,-}^\dag \ket{00}\bra{10}M_{-,+}\rho_k M_{-,+}^\dag \ket{10}}+\\
&\sqrt{\bra{11}M_{-,-}\rho_k M_{-,-}^\dag \ket{11}\bra{01}M_{-,+}\rho_k M_{-,+}^\dag \ket{01}}\\
&\qquad\qquad = \left(\tfrac{\NNN^+_{\sqrt{1-\eta}\alpha}\NNN^-_{\sqrt{1-\eta}\alpha}}{\NNN^+_{\alpha}\NNN^-_{\alpha}}\right)V^{\text{parity}}(\rho_k).
\end{align*}
Taking the expectation value of both sides we find the result of~\eqref{eq:Lyap}. {Thus $V^{\text{parity}}$ exponentially} decays to zero at a rate
$$
r_{\text{parity}}=\frac{1}{2}\log\left(\frac{1-e^{-4|\alpha|^2}}{1-e^{-4(1-\eta)|\alpha|^2}}\right).
$$
{In the case of a fully EP-QND measurement, 
{i.e.~assuming 
$\frac{\NNN_{\sqrt{\eta}\alpha}^+}{\NNN_\alpha^+} \simeq \frac{\NNN_{\sqrt{\eta}\alpha}^-}{\NNN_\alpha^-}$ and 
$\frac{\NNN_{\sqrt{\eta}\alpha}^+}{\NNN_\alpha^-} \simeq \frac{\NNN_{\sqrt{\eta}\alpha}^-}{\NNN_\alpha^+}$
in the Kraus operators,} the two terms in the Lyapunov function \eqref{eq:Lyapa} would decay at the same rate and $r_{\text{parity}}$ represents precisely the parity measurement strength. Indeed, in this case 
{ we would obtain the same result with the alternative Lyapunov function
$V^{\text{parity}}_{\text{ideal}}(\rho)= \frac{1}{2}\sqrt{1-P(\rho)^2} \;\; ,$
where $P(\rho)=\tr{\sigma_z\otimes\sigma_z \rho}$.}}

One can also analytically calculate the dephasing rate induced by the transmission loss. To this aim we define the coherence function
$$
{C(\rho)=\Big|\bra{00}\rho\ket{11}\Big|+\Big|\bra{01}\rho\ket{10}\Big|.}
$$ 
Similar calculations as above lead to
$$
\langle C(\rho_{k+1})\rangle=\frac{\sqrt{1-e^{-4\eta|\alpha|^2}}}{\sqrt{1-e^{-4|\alpha|^2}}}\langle C(\rho_{k})\rangle.
$$
Therefore the coherence function $C(\rho)$ {exponentially} decays to zero at a rate
$$
r_{\text{dephasing}}=\frac{1}{2}\log\left(\frac{1-e^{-4|\alpha|^2}}{1-e^{-4\eta|\alpha|^2}}\right).
$$
\begin{figure}
\includegraphics[height=60mm, trim=0mm 0mm 0mm 0mm, clip=true]{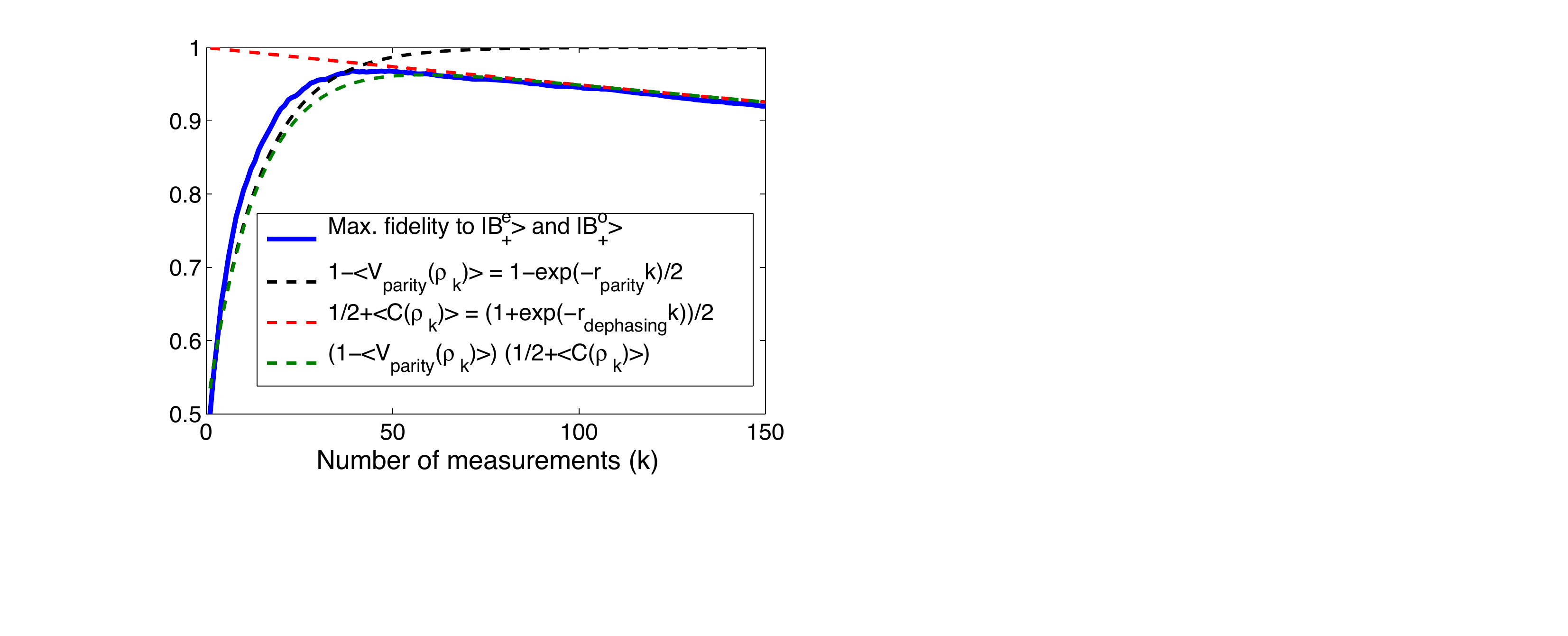}
\caption{{The performance of the near EP-QND parity measurement is explained by two rates. The measurement strength $r_{\text{parity}}$ indicates how fast the state is projected onto a parity eigenspace. The dephasing rate $r_{\text{dephasing}}$ indicates how fast the coherence inside a parity eigenspace vanishes. Here, we have fixed $\eta=.75$ and $|\alpha|^2=2$. The blue (solid) curve illustrates the maximum fidelity to one of the two Bell states $\ket{B_+^e}$ and $\ket{B_+^o}$, when the two qubits are initialized in  the state $(\ket{0}+\ket{1})_A(\ket{0}+\ket{1})_B/2$ and we take an average over 1000 Monte-Carlo trajectories. The black (dashed) curve illustrates the average parity converging at the characteristic rate of the Lyapunov function $V_{\text{parity}}(\rho)$. The red (dashed) curve, decreasing with the same characteristic rate as the coherence function $C(\rho)$, illustrates the dephasing inside a parity eigenspace. The green (dashed) curve represents the product of both these effects. This theoretical curve fits reasonably well with the simulations (blue solid curve).}  \label{fig:ThyVsSim}}
\end{figure}

Figure \ref{fig:ThyVsSim} provides a comparison of the simulations to the above analytical results. As can be seen, the performance of the parity measurement protocol can be well-explained using the above two rates. The slight mismatch between the raising rate of the blue curve (fidelity to Bell states) and the theoretical $r_{\text{parity}}$ can be explained by the nonlinear relation between the fidelity and the Lyapunov function $V_{\text{parity}}$. This nonlinearity makes it impossible to translate the rate  $r_{\text{parity}}$ into a precise exponential rate for the raising dynamics of the average fidelity.

{As explained through the main text, using these rates, it is possible to estimate the maximum achievable fidelity as a function of $\eta$ and $|\alpha|^2$. Indeed, this maximum fidelity is achieved, {approximately}, when the contribution of the first effect (parity projection) becomes equivalent to that of the second one (dephasing inside a parity eigenspace). This leads to the transcendental equation $e^{-r_{\text{parity}}T}+e^{-r_{\text{dephasing}}T}=1$. In the limit of $r_{\text{parity}}\gg r_{\text{dephasing}}$ (equivalent to $e^{4(2\eta-1)|\alpha|^2}\gg1$), one can approximately replace  $e^{-r_{\text{dephasing}}T}$ by $1-r_{\text{dephasing}}T$, and therefore the solution to the transcendental equation is well estimated by}
$$
T_{\text{meas}}=\frac{1}{r_{\text{parity}}}W_0\left(\frac{r_{\text{parity}}}{r_{\text{dephasing}}}\right),
$$
where $W_0$ is the Lambert W-function. The simulations of Fig.~3 in the main text illustrate this result.

\subsection*{Feedback stabilization of an entangled state}

{
As outlined in the main text, the EP-QND parity measurement allows to stabilize a highly entangled state through a simple feedback mechanism. We here provide some details about this scheme.

We first note that the feedback decision requires to know the respective parity populations. This knowledge is typically obtained by a quantum filter, i.e.~a computer estimating the state by simulating the evolution Eq.(3) of the main text associated to the respective detection results. Such filters are known to be stable \cite{rouchon2011fidelity}. 

In the present case, the quantum filter can be simplified significantly, namely by \emph{discarding all off-diagonal components in the Bell state basis.} Indeed, first note that the measurement does not depend on coherences among subspaces of different parity. Moreover, the measurement itself completely destroys, in a single iteration, any coherences between subspaces of different parity. The $\pi/2$ pulses ``export'' coherences possibly present between e.g.~$\ket{B^e_+}$ and $\ket{B^e_-}$ into coherences between $\ket{B^e_+}$ and $\ket{B^o_+}$, i.e.~between different parity subspaces, which are thus destroyed at the next measurement; while they ``import'' into e.g.~the even parity eigenspace, the coherences previously present between $\ket{B^e_+}$ and $\ket{B^o_+}$, which are none since those two Bell states belong to different parity eigenspaces and their coherences were thus destroyed by the last measurement. Thus after one initial weak measurement step at most, no relevant coherences among Bell states will survive. This allows to update just the \emph{populations} on the four Bell states, as in a classical filter for a partially observed Markov chain. This filter, updating the populations on the four Bell states, is the only computation required for the feedback: the action itself just requires a binary decision, to switch towards the predominantly populated parity.\vspace{3mm}

To analyze the feedback more explicitly, we note that instead of applying the $\pi/2$ pulses to rotate the \emph{state} in the Schr\"odinger picture, we can reformulate the dynamics by applying the $\pi/2$ pulses to the \emph{measurement} scheme, in an equivalent Heisenberg picture. The corresponding measurements then alternate between the parity measurement $\sigma_z \otimes \sigma_z$ in $z$-basis, and a parity measurement $\sigma_x \otimes \sigma_x$ in $x$-basis. The corresponding state is subject to $\pi$-pulses in the respective $z$ or $x$ basis. A $\pi$ pulse in $z$ basis (resp.~$x$ basis) does not change the populations of $x$-parity subspaces (resp.~of $z$-parity subspaces), and is applied to increase the population in $\text{span}\{\ket{B^e_+},\ket{B^e_-}\}$ (resp.~$\text{span}\{\ket{B^e_+},\ket{B^o_+}\}$). The stochastic convergence of the measured system towards a definite parity in both $x$ and $z$ coordinates ensures that, in absence of other effects, the population in $\text{span}\{\ket{B^e_+},\ket{B^e_-}\} \cap \text{span}\{\ket{B^e_+},\ket{B^o_+}\} = \ket{B^e_+}$ would increase in expectation until reaching 100\%. The dephasing effect limits the actual fidelity for a given $\alpha$. Taking $\alpha$ larger allows to increase the fidelity, but slows down the convergence, such that a tradeoff value of $\alpha$ must be selected with respect to other decoherence effects acting on the system.

We can quantify the performance of this feedback scheme using the characteristic convergence rates computed in the previous section. We here keep the viewpoint of the previous paragraph, of performing the measurement alternatively in $z$-basis and in $x$-basis. Then a similar continuous-time model for the populations $p_k$ of Bell state $\ket{k}$ writes:
\begin{eqnarray}\label{eq:rprd}
\tfrac{d}{dt}p_{B^e_+} & = & r_{\text{parity}}(\frac{p_{B^e_-}}{2}+ \frac{p_{B^o_+}}{2}) 
 - \frac{r_{\text{dephasing}}}{2}(p_{B^e_+} - \frac{p_{B^e_-}}{2} - \frac{p_{B^o_+}}{2}) \\ \nonumber
\tfrac{d}{dt}p_{B^e_-} & = & r_{\text{parity}}(\frac{p_{B^o_-}}{2}- \frac{p_{B^e_-}}{2}) 
 - \frac{r_{\text{dephasing}}}{2}(p_{B^e_-} - \frac{p_{B^e_+}}{2} - \frac{p_{B^o_-}}{2}) \\ \nonumber
\tfrac{d}{dt}p_{B^o_+} & = & r_{\text{parity}}(\frac{p_{B^o_-}}{2}- \frac{p_{B^o_+}}{2}) 
 - \frac{r_{\text{dephasing}}}{2}(p_{B^o_+} - \frac{p_{B^e_+}}{2} - \frac{p_{B^o_-}}{2}) \\ \nonumber
\tfrac{d}{dt}p_{B^o_-} & = & - r_{\text{parity}}p_{B^o_-} 
 - \frac{r_{\text{dephasing}}}{2}(p_{B^o_-} - \frac{p_{B^e_-}}{2} - \frac{p_{B^o_+}}{2}) \, .
\end{eqnarray}
To obtain these equations, we first use the fact that the measurement destroys any coherences among Bell states, so only populations are relevant. Then we note that, assuming an ideal parity measurement, one measurement iteration out of two drives the system at a rate $r_{\text{parity}}$ from even to odd parity in $z$-basis while maintaining the parity in $x$-basis, i.e.~sending population from $\ket{B^o_+}$ to $\ket{B^e_+}$ and from $\ket{B^o_-}$ to $\ket{B^e_-}$ at an overall rate $r_{\text{parity}}/2$; and one measurement iteration out of two drives the system at a rate $r_{\text{parity}}$ from even to odd parity in $x$-basis while maintaining the parity in $z$-basis, i.e.~sending population from $\ket{B^e_-}$ to $\ket{B^e_+}$ and from $\ket{B^o_-}$ to $\ket{B^o_+}$. Similarly, for one measurement iteration ouf of two, the channel imperfection implies dephasing inside the even parity manifold in $z$-basis; and for the other iteration in $x$-basis. The loss of coherence $\bra{00}\rho\ket{11}$ at a rate $r_{\text{dephasing}}$ corresponds to a phase flip at rate $r_{\text{dephasing}}/2$. Thus the e.g.~$z$-basis measurement implies an exchange of populations between $\ket{B^e_-}$ and $\ket{B^e_+}$, and between $\ket{B^o_-}$ and $\ket{B^o_+}$, at an overall rate $r_{\text{dephasing}}/4$. 

The nonlinear relation between $V^{\text{parity}}$ and the fidelities makes it impossible to be more accurate than order of magnitude about $r_{\text{parity}}$, and we have noted in simulations that the true dynamics is more complicated than the linear model \eqref{eq:rprd}. Moreover, clearly this continuous-time approximation is only valid for $r_{\text{parity}},r_D \ll 1/T_{\text{meas}}$. However, the model \eqref{eq:rprd} gives valuable indications about the dependence of the whole scheme on $\alpha$ (through $r_{\text{parity}}$ and $r_{\text{dephasing}}$) and allows to efficiently optimize $\alpha$ for given settings.

The steady state of \eqref{eq:rprd} is obtained with
$$p_{B^e_+} = \frac{\delta^2}{(1+\delta)^2} \;\;\; \text{where } \delta = \frac{r_{\text{parity}}+r_{\text{dephasing}}/2}{r_{\text{dephasing}}/2} \, .$$
Again, there is the tradeoff that larger $\alpha$ implies larger $p_{B^e_+}$ at steady state, but slower convergence towards this steady state.

To practically illustrate this tradeoff, we can add to the picture the qubit relaxation:
\begin{eqnarray}\label{eq:decoh}
\tfrac{d}{dt}\rho & = & \sum_{k\in \{A,B\}} L_k \rho L_k^\dag - \tfrac{1}{2}(L_k^\dag L_k \rho + \rho L_k^\dag L_k) \\
\nonumber & & \text{with } L_k = \sqrt{\frac{T_{\text{meas}}}{T_1}} \ket{0}\bra{1}_k \, .
\end{eqnarray}
Here $\ket{0}\bra{1}_k$ is the qubit lowering operator, $T_1$ is the relaxation time and $T_{\text{meas}}$ the time taken by one measurement iteration. One checks that in the Bell state basis, when discarding off-diagonal terms in $\rho$, the effect of this qubit decoherence is equivalent to changing $r_{\text{dephasing}}/2$ to $r_{\text{dephasing}}/2 + \frac{T_{\text{meas}}}{T_1}$ in \eqref{eq:rprd}. Thus in good approximation, for $r_{\text{parity}} \gg r_{\text{dephasing}}, \, \frac{1}{T_1}$, the optimal choice of $\alpha$ is the one maximizing
\begin{equation}\label{eq:maxpBe}
p_{B^e_+} = \frac{\delta^2}{(1+\delta)^2} \;\;\; \text{with } \delta = 1 + \frac{r_{\text{parity}}}{\frac{r_{\text{dephasing}}}{2}+\frac{T_{\text{meas}}}{T_1}} \, .
\end{equation}
This is equivalent to maximizing 
$$f(\alpha) := \frac{r_{\text{parity}}(\alpha)}{r_{\text{dephasing}}(\alpha)/2 + T_{\text{meas}}/T_1} \, .$$
When $\alpha$ increases, $\frac{r_{\text{parity}}}{r_{\text{dephasing}}}$ increases but both $r_{\text{parity}}$ and $r_{\text{dephasing}}$ decrease, so at some point the constant $T_{\text{meas}}/T_1$ starts making it disadvantageous to further increase $\alpha$. The optimum can easily be computed numerically. 

We have simulated the actual system dynamics, i.e.~adding qubit decoherence \eqref{eq:decoh} to Eq.(3) of the main text, for different values of $\eta$ and $T_{\text{meas}}/T_1$. For $\alpha$ we have taken the value that maximizes \eqref{eq:maxpBe}, as well as a few values close to it in order to confirm that we hit close to the actual optimum fidelity. For each set of parameters, we have performed 5000 Monte-Carlo simulations in order to estimate the achieved fidelity. Figure \ref{fig:fbfid} shows these simulation results. The value of $\alpha$ obtained by maximizing \eqref{eq:maxpBe} appears to indeed be (close to) optimal, while the actual value returned by the formula \eqref{eq:maxpBe} slightly overestimates the achieved fidelity. The results show that if e.g.~3000 measurements can be performed during a qubit lifetime (comparing 100$\mu$s lifetime to 100ns duration of the most demanding measurement operation) and transmission fidelity reaches about 85\%, then the steady-state entanglement fidelity can be pushed up to 99\%.
}

\begin{figure}
\includegraphics[width=100mm, trim=10mm 0mm 20mm 5mm, clip=true]{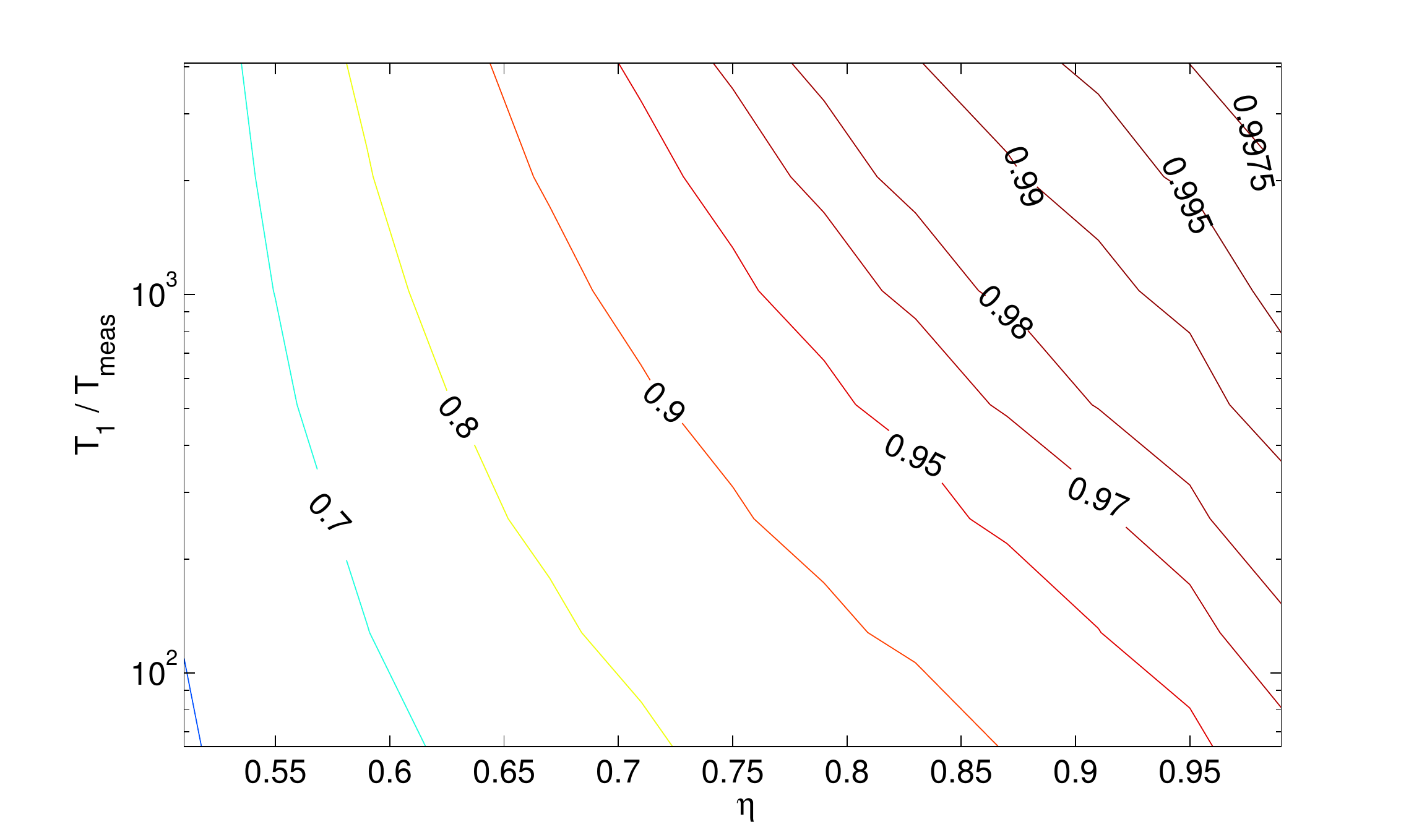}
\caption{Fidelities to the Bell state $\ket{B^e_+}=(\q{00}+\q{11})/\sqrt{2}$ obtained by our remote entanglement feedback stabilization scheme in presence of individual qubit decay \eqref{eq:decoh}. For each value of $\eta$ (transmission fidelity of the meter quantum channel for the remote parity measurement) and of $T_1/T_{\text{meas}}$ (qubit characteristic lifetime expressed in number of measurement iterations), we have estimated the optimal value of $\alpha$ by maximizing \eqref{eq:maxpBe}.
We have then run 5000 Monte-Carlo simulations with values of $\alpha$ up to 2 times larger or smaller than this estimated optimum and selected the best result. The latter turned out to be the theoretically computed $\alpha$, although the corresponding maximal fidelity is slightly overestimated by \eqref{eq:maxpBe}. To estimate steady-state fidelity, we have averaged the fidelity obtained between measurement iterations $3 \, T_1/T_{\text{meas}}$ and $6 \, T_1/T_{\text{meas}}$, when starting from $\ket{B^e_+}$. (Here fidelity to $\q{\psi_0}$ means $\qd{\psi_0} \rho \q{\psi_0}$.)
}\label{fig:fbfid}
\end{figure}

\subsection*{Remote entanglement and EP-QND parity measurement are essentially equivalent}

We have just shown how an EP-QND parity measurement is sufficient to enable remote preparation of a perfectly entangled state through local corrective actions and classical communication. Then the well-known fact that entanglement can impossibly be increased with only local actions and classical communication, implies that EP-QND parity measurement necessarily requires communication over a quantum channel; this justifies the general setup of Fig.1a of the main text.

The converse is also true, making these two resources essentially equivalent: owning a remote perfectly entangled state allows to perform an EP-QND parity measurement via just local actions and classical communication. Indeed assume that we have auxiliary qubits $\q{q_{A1}}$ and $\q{q_{B1}}$ located at $A$ and $B$ respectively and initially prepared in the state $(\q{00}+\q{11})_{A1,B1}\, /\sqrt{2}$. Applying a local CNOT gate on $\q{q_{A1}}$ conditioned by $\q{q_{A}}$ and on $\q{q_{B1}}$ conditioned by $\q{q_{B}}$, we get for an initial even-parity state $(\q{00}\pm\q{11})_{A,B}$:
$$(\q{00}+\q{11})_{A1,B1}\, (\q{00}\pm\q{11})_{A,B} \, ;$$
and for an initial odd-parity state $(\q{01}\pm\q{10})_{A,B}$:
$$(\q{01}+\q{10})_{A1,B1}\, (\q{01}\pm\q{10})_{A,B} \; .$$
Thus the initial state of the target qubits remains unaffected (EP-QND property), and when measuring the auxiliary qubits in their local canonical basis $\q{0},\q{1}$ the correlation between detection results shall give an indication about the parity; this information shall be perfectly discriminating if the detection is perfect, while if the detection is imperfect it will give only partial information without affecting the EP-QND character.

\end{document}